\documentclass[12pt]{iopart}
 \usepackage{epsfig}

\def\gsim{\,\raise0.3ex\hbox{$>$\kern-0.75em\raise-1.1ex\hbox{$\sim$}}\,}
\def\lsim{\,\raise0.3ex\hbox{$<$\kern-0.75em\raise-1.1ex\hbox{$\sim$}}\,}

\usepackage{iopams}  
\usepackage{pslatex}
\usepackage{txfonts}

\newcommand{\sqrtsnn}{\sqrt{s_{_{NN}}}}

\providecommand{\mean}[1]{\ensuremath{\left<#1\right>}}

\providecommand{\dNdeta}{dN_{\rm ch}/d\eta|_{\eta=0}}

\begin{document}

\title{Identified hadron spectra in Pb-Pb at $\sqrtsnn$ = 5.5 TeV: hydrodynamics+pQCD predictions}
\vspace{-2mm}
 
\author{Fran\c{c}ois Arleo$\hspace{1mm}^{1,2}$, David d'Enterria$\hspace{1mm}^2$, Dmitri Peressounko$\hspace{1mm}^3$}
\address{{\bf 1.} LAPTH, UMNR CNRS/Univ. Savoie, B.P. 110, 74941 Annecy-le-Vieux Cedex}
\address{{\bf 2.} CERN/PH, CH-1211 Geneva 23}
\address{{\bf 3.} RRC ``Kurchatov Institute'', Kurchatov Sq. 1, Moscow 123182}

\begin{abstract}
  The single inclusive charged hadron $p_T$ spectra in Pb-Pb collisions at the LHC, 
  predicted by a combined hydrodynamics+perturbative QCD (pQCD) approach are presented.
\end{abstract}

\noindent
We present predictions for the inclusive transverse momentum distributions of pions, kaons and 
(anti)protons produced at mid-rapidity in Pb-Pb collisions at $\sqrtsnn$ = 5.5 TeV based on 
hydrodynamics+pQCD calculations. The bulk of the spectra ($p_T\lesssim$ 5 GeV/$c$) in central 
Pb-Pb at the LHC is computed with a hydrodynamical model -- successfully tested at RHIC~\cite{d'Enterria:2005vz} --
using an initial entropy density extrapolated empirically from the hadron multiplicities 
measured at RHIC: $\dNdeta/(0.5\,N_{\rm part})\approx 0.75\ln(\sqrtsnn/1.5)$~\cite{Adler:2004zn}. 
Above $p_T\approx$ 3 GeV/$c$, additional hadron production from (mini)jet fragmentation 
is computed from collinearly factorized pQCD cross sections at next-to-leading-order (NLO)
accuracy~\cite{Aurenche:1999nz}. We use recent parton distribution functions (PDF)~\cite{Pumplin:2002vw} 
and fragmentation functions (FF)~\cite{Albino:2005me}, modified respectively
to account for initial-state shadowing~\cite{deFlorian:2003qf} 
and final-state parton energy loss~\cite{Arleo:2007bg}.\\
\vspace{-2mm}

\noindent
We use cylindrically symmetric boost-invariant 2+1-D relativistic hydrodynamics, fixing the initial
conditions for Pb-Pb at $b$ = 0 fm and employing a simple Glauber prescription to obtain the 
corresponding source profiles at all other centralities~\cite{d'Enterria:2005vz}. The initial source is 
assumed to be formed at a time $\tau_0 = 1/Q_s\approx$ 0.1 fm/$c$, with an initial entropy density 
of $s_0$~=~1120~fm$^{-3}$ (i.e. 
$\varepsilon_o\propto s_0^{4/3}\approx$ 650~GeV/fm$^3$) so as to reproduce the 
expected final hadron multiplicity $\dNdeta\approx$ 1300 at the LHC~\cite{Adler:2004zn}.
We follow the evolution of the system by solving the equations of ideal hydrodynamics 
including the current conservation for net-baryon number (the system is almost baryon-free, 
$\mu_B\approx$ 5 MeV). For temperatures above (below) $T_{\rm crit}\approx$ 170 MeV 
the system is described with a QGP (hadron gas) equation of state (EoS).
The QGP EoS -- obtained from a parametrization to recent lattice QCD results -- is Maxwell
connected to the hadron resonance gas phase assuming a first-order phase transition.
As done for RHIC energies, we chemically freeze-out the system (i.e. fix the hadron ratios) at $T_{\rm crit}$.
Final state hadron spectra are obtained with the Cooper-Frye prescription 
at $T_{\rm fo}\approx$ 120 MeV followed by decays of unstable resonances using the known 
branching ratios. Details can be found at~\cite{d'Enterria:2005vz}.\\
\vspace{-2mm}

\noindent
Our NLO pQCD predictions are obtained with the code of ref.~\cite{Aurenche:1999nz} 
with 
all scales set to $\mu=p_{T}$. Pb-Pb yields are obtained scaling the NLO 
cross-sections by the number of incoherent nucleon-nucleon collisions for each centrality class given by
a Glauber model ($N_{\rm coll}$ = 1670, 12.9 for 0-10\%-central and 60-90\%-peripheral). 
Nuclear (isospin and shadowing) corrections of the CTEQ6.5M PDFs~\cite{Pumplin:2002vw} 
are introduced using the NLO nDSg 
parametrization~\cite{deFlorian:2003qf}. Final-state energy loss in the hot and dense
medium is accounted for by modifying the AKK FFs~\cite{Albino:2005me} with BDMPS
quenching weights as described in~\cite{Arleo:2007bg}. The BDMPS medium-induced 
gluon spectrum depends on a single scale $\omega_c=\mean{\hat{q}}\,L^2$, 
related to the transport coefficient and length of the medium. We use 
$\omega_c\approx$ 50 GeV, from the expected energy dependence of the 
quenching parameter and the measured $\omega_c\approx$ 20 GeV at 
RHIC~\cite{Arleo:2007bg}. The inclusive hadron spectra in central Pb-Pb 
are suppressed by up to a factor $\sim$10 (2), $R_{PbPb}\approx$~0.1 (0.5), at $p_T$ = 10 (100) GeV/$c$.\\
\vspace{-2mm}

\begin{figure}[htbp]
\begin{centering}
\hspace{6mm}
\includegraphics[width=0.45\textwidth]{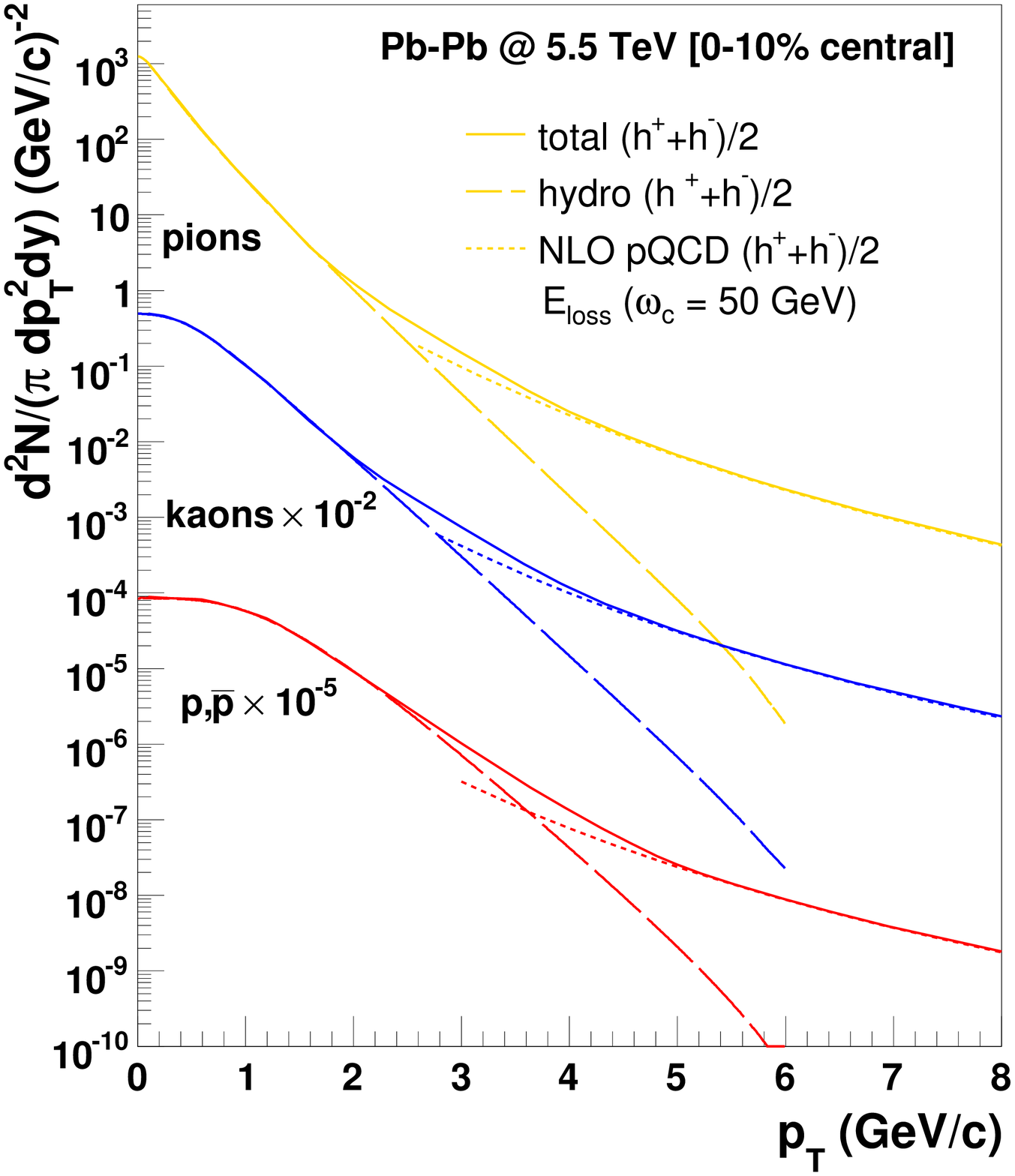}
\hspace{6mm}
\includegraphics[width=0.45\textwidth]{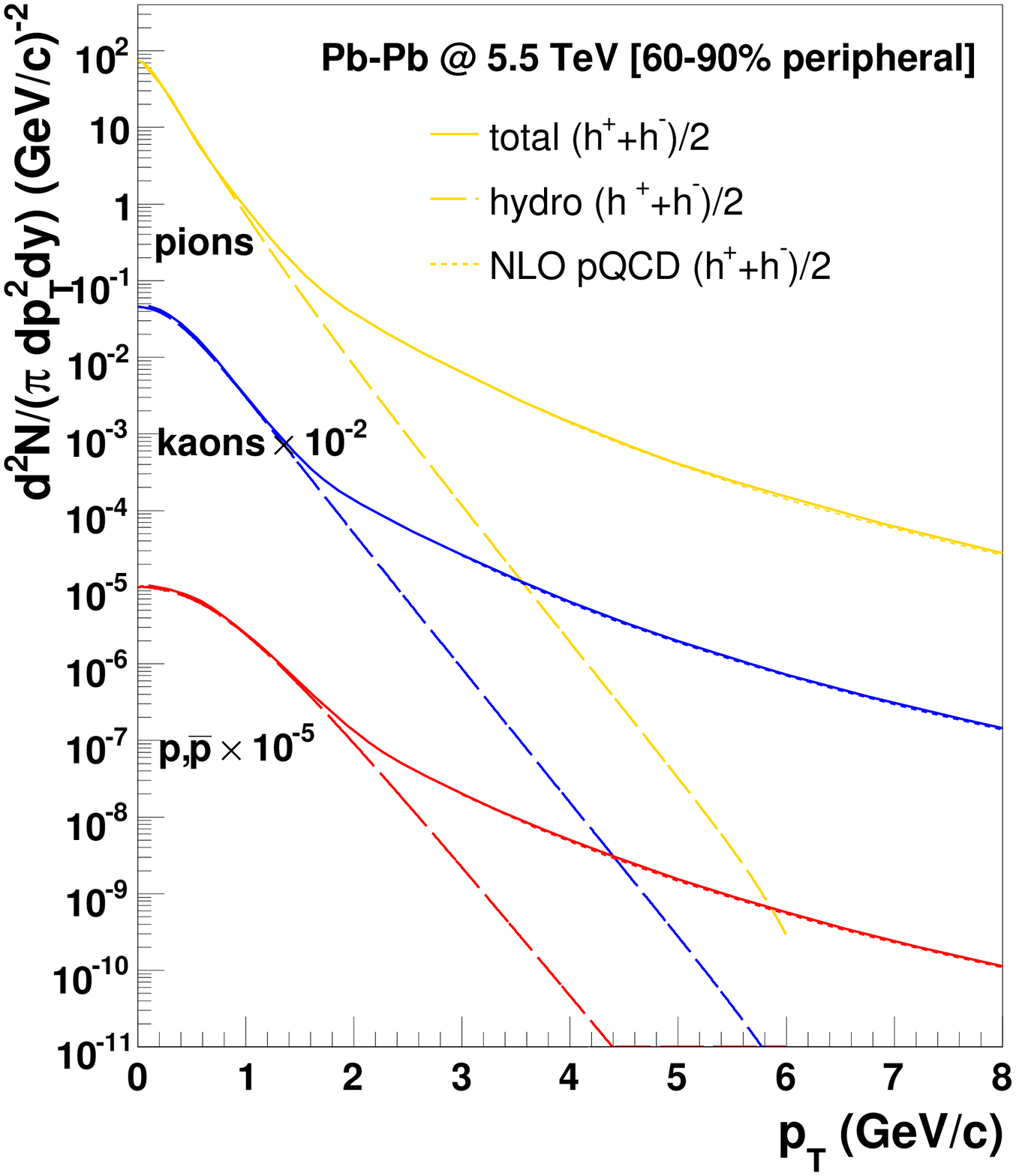}
\end{centering}
\vspace{-1cm}
\caption{\label{fig:spectra}Spectra at y=0 for $\pi^{\pm}$,$K^{\pm}$, $p,\bar{p}$ in 0-10\% central (left)
and 60-70\% peripheral (right) Pb-Pb at $\sqrtsnn$ = 5.5 TeV, obtained with hydrodynamics + (quenched) pQCD calculations.}
\end{figure}

\noindent
Our predictions for the identified hadron spectra in Pb-Pb collisions at 5.5 TeV are shown 
in Figure~\ref{fig:spectra}. The hydrodynamical contribution dominates over the (quenched) pQCD 
one up to $p_T\approx$ 4 (1.5) GeV/$c$ in central (peripheral) Pb-Pb. As expected, the hydro-pQCD 
$p_T$ crossing point increases with the hadron mass. In the absence of recombination effects 
(not included here), bulk protons may be boosted up to $p_T\approx$ 5 GeV/$c$ in central Pb-Pb at the LHC.

\vspace{-2mm}

\vspace{-3mm}
\noindent
Dd'E. and D.P. acknowledge resp. support from 6th EU FP contract MEIF-CT-2005-025073 and
MPN Russian Federation grant NS-1885.2003.2.



\section*{References}
\begin{thebibliography}{10}
\vspace{-2mm}

\bibitem{d'Enterria:2005vz}
  D.~d'Enterria and D.~Peressounko,
  Eur.\ Phys.\ J.\  C {\bf 46}, 451 (2006)
  [arXiv:nucl-th/0503054] and refs. therein.

\bibitem{Adler:2004zn}
  S.~S.~Adler {\it et al.}  [PHENIX Collaboration],
  Phys.\ Rev.\  C {\bf 71} (2005) 034908
  [arXiv:nucl-ex/0409015].

\bibitem{Aurenche:1999nz}
  P.~Aurenche {\it et al}, 
  Eur.\ Phys.\ J.\  C {\bf 13}, 347 (2000)
  [arXiv:hep-ph/9910252].

\bibitem{Pumplin:2002vw}
  J.~Pumplin {\it et al.}, 
  JHEP {\bf 0207} (2002) 012
  [arXiv:hep-ph/0201195].

\bibitem{Albino:2005me}
  S.~Albino, B.~A.~Kniehl and G.~Kramer,
  Nucl.\ Phys.\  B {\bf 725}, 181 (2005)
  [arXiv:hep-ph/0502188].

\bibitem{deFlorian:2003qf}
  D.~de Florian and R.~Sassot,
  Phys.\ Rev.\  D {\bf 69}, 074028 (2004)
  [arXiv:hep-ph/0311227].

\bibitem{Arleo:2007bg}
  F.~Arleo, JHEP 07 (2007) 032 
  arXiv:0706.1848 [hep-ph], and  arXiv:0707.2320 [hep-ph].


\end{thebibliography}
\end{document}